# A class of Šidák-type tests based on maximal precedence and exceedance statistic

Niladri Chakraborty[a]*, Di Cui[b], Min Xie[b,c]

[a]*Department of Mathematical Statistics and Actuarial Science, University of the Free State, Bloemfontein, South Africa;* [b]*Department of Systems Engineering and Engineering Management, City University of Hong Kong, Hong Kong SAR;* [c]*City University of Hong Kong Shenzhen Research Institute, Nanshan, Shenzhen, Guangdong, P. R. China*

**Abstract:** A class of nonparametric two-sample tests has been proposed in this article. As a generalization of the original Šidák's test, the proposed test statistic is developed as the sum of the maximal precedence and maximal exceedance statistics. Unlike the Šidák-type precedence-exceedance test and the maximal precedence test, the proposed test is suitable for a two-sided alternative while being free from any parametric assumption. Exact distribution of the test statistic is obtained under the null as well as under the Lehmann alternative. Power value comparison has been carried out that shows the competency of the proposed test as a useful alternative to a number of existing tests based on precedence-exceedance statistics. Real-life example is provided to illustrate the application of the proposed test.

1. Introduction

Suppose $X_1, X_2, X_3, \ldots, X_m$ and $Y_1, Y_2, Y_3, \ldots, Y_n$ are random samples from univariate, absolutely continuous distributions $F$ and $G$, respectively. A two-sample life testing problem is given by

$$H_0: F = G \text{ ag. } H_1: F > G. \tag{1}$$

As one of the pioneering contributions in distribution-free testing of $H_0$ against $H_1$, Šidák projected a test based on the sum of preceding $X$ samples and exceeding $Y$ samples (Seidler (2000)). The null distribution of this sum test and corresponding critical values were obtained by Šidák and Vondracek (1957).

Counting against the extreme order statistics may be influenced by the presence of outliers in the sample. To curtail such influence, use of the $r^{th}$ order statistics is well accepted in literature; see, for example, van der Laan and Chakraborti (2001), Balakrishnan and Ng (2006). These tests are known as precedence tests that are based on the counts of $X$ sample preceding $Y_{(r)}$, the $r^{th}$ order statistic from $Y$ sample. Generalization of precedence tests based on record statistic and censored data have been proposed by Balakrishnan et al. (2008a, 2008b). A comprehensive overview on precedence tests is provided in Balakrishnan and Ng (2006).

Stoimenova and Balakrishnan (2011) proposed a class of tests based on the minimum of preceding and exceeding counts with respect to the $r^{th}$ order statistics from both samples. A Šidák-type sum test based on the preceding and exceeding counts has been proposed by Stoimenova and Balakrishnan (2017). In precedence/exceedance tests, total counts of one sample with respect to the $r^{th}$ order statistic from the other sample is considered. This might lead to a 'masking effect' as noted by Balakrishnan and Frattina (2000).

For example, in a life-testing experiment with two groups of size $n = 15$, the critical limit is supposed to be $c = 9$ for $r = 4$ in a standard precedence test (van der Laan and Chakraborti (2001)). This means, the test will not reject $H_0$ when there are 8 observations from the $X$ sample before $Y_{(4)}$. However, this decision might be 'masked' in the sense that it does not consider the frequency distribution of $X$ values among $Y$ values. If all 8 observations fall below $Y_{(1)}$ or between $Y_{(1)}$ and $Y_{(2)}$, there might be a shift in the distribution of $Y$ sample.

Balakrishnan and Frattina (2000) noted that the 'masking effect' can reduce the performance of precedence test. To ease such masking effect, they proposed a maximal precedence test. This test is based on the maximal frequency of failures from one sample occurring between every consecutive ordered failures, up to the 2$^{nd}$ ordered failure, from the other sample. Balakrishnan and Ng (2001) extended the maximal precedence test up to the $r^{th}$ order statistic. Generalizing the precedence and maximal precedence tests, Ng and Balakrishnan (2005) proposed weighted precedence and weighted maximal precedence tests. Other relevant developments based on precedence-exceedance statistics include Ng et al. (2007), Ng and Balakrishnan (2010), Balakrishnan et al. (2010), Graham et al. (2014), Balakrishnan et al. (2015), Erem and Bayramoglu (2017), Graham et al. (2017), Mukherjee et al. (2018), Chakraborty et al. (2018), Li et al. (2019), Balakrishnan et al. (2020), Erem (2020) among others.

In this article, we propose a Šidák-type test based on the maximal precedence and exceedance statistic. Precedence/exceedance type tests are useful when a quick, reliable decision is needed with few early failure data, especially when expensive units are put on a life-test (Ng and Balakrishnan, 2005). We obtain the joint distribution of maximal precedence and maximal exceedance statistics. Using the joint distribution, an exact distribution of the Sidak-type statistic is obtained under the null and Lehmann alternative.

Rest of the article is organized as follows. A formal description of the proposed test and an example on practical implementation is provided in Section 2. In Section 3, exact null distribution of the test statistic and the critical values are obtained. Large sample approximation of the null distribution is also provided. In Section 4, exact distribution of the test statistic under the Lehmann alternative and corresponding power values are obtained. Then we compare the power of the proposed test with other precedence/exceedance tests under the Lehmann alternatives and also under different lifetime distributions. Some concluding remarks are presented in Section 5. Formal proofs of the theorems are consigned to the Appendix.

## 2. Description of the proposed test

Let us call $X$ sample as the *training* group and $Y$ sample as the *test* group, without loss of generality. Order the test group and denote $Y_{(i)}$ the ordered $Y$ observations. Define

$f_{pi}$ = Number of $X$ values falling in the interval $(Y_{(i-1)}, Y_{(i)}]$ for $i = 1, 2, 3, \ldots, r$ with $Y_{(0)} = 0$,

$f_{ei}$ = Number of $X$ values falling in the interval $[Y_{(n-s+i)}, Y_{(n-s+1+i)})$ for $i = 1, 2, 3, \ldots, s$ with $Y_{(n+1)} = \infty$.

The maximal precedence and exceedance statistics, denoted by $P_r$ and $E_s$ respectively, is defined as

$P_r = \max\{f_{pi}; i = 1, 2, 3, \ldots, r\}$,

$E_s = \max\{f_{ei}; i = 1, 2, 3, \ldots, s\}$.

Let us define a Šidák-type test statistic based on maximal precedence and maximal exceedances as

$$T_{r,s} = P_r + E_s, \qquad (2)$$

where, $r = [\rho_1 n] + 1$, $s = [\rho_2 n] + 1$, $0 \leq \rho_1, \rho_2 < 1$, $2 \leq r + s \leq n$. When $\rho_1 = \rho_2 = \rho$, we have $r = s = [\rho n]+1$, $0 \leq \rho < 0.5$. Then the test statistic in (2) is given by

$$T_r = P_r + E_r.  \qquad (3)$$

The test defined by $T_r$ includes Šidák's test as a special case when $r = 1$.

The following example is useful in numerical illustration of the proposed $T_{r,s}$-test. The data set is obtained from Lawless (2011, Example 5.4.3, p. 240). Two types of electrical cable insulation were subjected to increasing voltage stress in a laboratory test. Consider $X$ and $Y$ to be the Type I and Type II insulation, respectively. Voltage levels (in kilovolts per millimeter) at which failures occurred were recorded.

**Example 2.1**. 20 specimens from groups $X$ and $Y$ were placed on a life-testing experiment. Failure voltages are presented in Table 1.

**Table 1. Failure voltages for two types of electrical cable insulation**.

| No. | X | Y | No. | X | Y |
|---|---|---|---|---|---|
| 1 | 32.0 | 39.4 | 11 | 46.5 | 57.1 |
| 2 | 35.4 | 45.3 | 12 | 46.8 | 57.2 |
| 3 | 36.2 | 49.2 | 13 | 47.3 | 57.5 |
| 4 | 39.8 | 49.4 | 14 | 47.3 | 59.2 |
| 5 | 41.2 | 51.3 | 15 | 47.6 | 61.0 |
| 6 | 43.3 | 52.0 | 16 | 49.2 | 62.4 |
| 7 | 45.5 | 53.2 | 17 | 50.4 | 63.8 |
| 8 | 46.0 | 53.2 | 18 | 50.9 | 64.3 |
| 9 | 46.2 | 54.9 | 19 | 52.4 | 67.3 |
| 10 | 46.4 | 55.5 | 20 | 56.3 | 67.7 |

In this experiment, we have $m = n = 20$ for each group. For $r, s = 1, 2, 3$, the values of $P_r$ and $E_s$ are presented in Table 2.

**Table 2. Computation of maximal precedence and exceedance statistics.**

| $r$ | $P_r$ | $s$ | $E_s$ |
| --- | --- | --- | --- |
| 1 | 3 | 1 | 0 |
| 2 | 3 | 2 | 0 |
| 3 | 10 | 3 | 0 |

3. **Null distribution of the test statistic**

We derive the exact null distribution of the test statistic $T_{r,s}$ defined in Eq. (2) and obtain the critical values for small sample sizes. A large sample approximation of the null distribution is obtained subsequently.

**3.1. Exact null distribution**

Given the joint distribution of $P_r$ and $E_s$ under $H_0$, the distribution function of $T_{r,s}$ can be obtained as

$$F_{r,s}(t|H_0) = P[T_{r,s} \leq t \,|H_0] = \sum_{k=0}^{t} \sum_{i=0}^{k} P[P_r = i, E_s = k - i \mid H_0], \quad 0 \leq t \leq m. \quad (4)$$

To derive the joint distribution of $P_r$ and $E_s$ under $H_0$, we obtain the distribution of the frequencies of $X_1, X_2, X_3, \ldots, X_m$ between each consecutive $Y_{(i)}$, before $Y_{(r)}$ and after $Y_{(s)}$. For $0 \leq f_{pi}, f_{ei} \leq m$, the joint probability mass function of $P_r$ and $E_s$ is given by

$P[P_r = i, E_s = j \mid H_0]$

$$= \sum_{\substack{f_{p1},f_{p2},\ldots,f_{pr},\ldots,f_{e1},f_{e2},\ldots,f_{es} \\ P_r=i\ ;\ E_s=j}} P[f_{p1},f_{p2},\ldots,f_{pr},\ f_{e1},f_{e2},\ldots,f_{es}\ |\ H_0] \qquad (5)$$

Next, we obtain the joint probability mass function $P[f_{p1},f_{p2},\ldots,f_{pr},\ f_{e1},f_{e2},\ldots,f_{es}\ |\ H_0]$.

**Theorem 3.1.** For $0 \leq f_{pi}, f_{ei} \leq m$, the joint probability mass function of $\{f_{p1},f_{p2},\ldots,f_{pr},f_{e1},f_{e2},\ldots,f_{es}\}$ under $H_0$ is given by

$$P[f_{p1},f_{p2},\ldots,f_{pr},\ f_{e1},f_{e2},\ldots,f_{es}\ |\ H_0] = \frac{\binom{m-\sum_1^s f_{ei}-\sum_1^r f_{pi}+n-r-s}{n-r-s}}{\binom{m+n}{n}},$$

$$= 0\ ,\ \text{otherwise.}$$

where $0 \leq \sum_1^s f_{ei} + \sum_1^r f_{pi} \leq m,\ r+s = 2,3,\ldots,n$.

*Proof.* The proof is presented in Appendix A2.

Replacing $P[f_{p1},f_{p2},\ldots,f_{pr},\ f_{e1},f_{e2},\ldots,f_{es}\ |\ H_0]$ from Theorem 3.1 in Eq. (5), and $j = (k-i)$, we get

$$P[P_r = i, E_s = k-i] = \sum_{\substack{f_{p1},f_{p2},\ldots,f_{pr},\ldots,f_{e1},f_{e2},\ldots,f_{es} \\ P_r=i\ ;\ E_s=k-i}} \frac{\binom{m-\sum_1^s f_{ei}-\sum_1^r f_{pi}+n-r-s}{n-r-s}}{\binom{m+n}{n}} \qquad (6)$$

Let us denote $N_1 = \sum_1^r f_{pi}$, $N_2 = \sum_1^s f_{ei}$, and $N = N_1 + N_2$. Note that, $0 \leq i \leq k \leq N \leq m$, $N_1 \leq rP_r$ and $N_2 \leq sE_s$. For all $\{f_{p1},f_{p2},\ldots,f_{pr},f_{e1},f_{e2},\ldots,f_{es}\}$, $P[f_{p1},f_{p2},\ldots,f_{pr},f_{e1},f_{e2},\ldots,f_{es}\ |\ H_0] = \frac{\binom{m-N+n-r-s}{n-r-s}}{\binom{m+n}{n}}$ will be same as long as $N_1 + N_2 = N$ is constant for different $\{f_{p1},f_{p2},\ldots,f_{pr},f_{e1},f_{e2},\ldots,f_{es}\}$. Using this property, we obtain an exact formula for $P[P_r = i, E_s = k-i]$ in Eq. (6) and subsequently obtain the null distribution of $T_{r,s}$ from Eq. (4).

### 3.2. Critical value and limiting distribution

Under $H_0$, the critical region for $T_{r,s}$ for a pre-specified significance level $\alpha$ is given by $\{c, (c+1), (c+2),\ldots, m\}$. The null hypothesis $H_0$ is rejected at level $\alpha$ if $T_{r,s} \geq c$, where $c$ is determined as the minimal value of all $c^*$ such that $P[T_{r,s} \geq c^*] \leq \alpha$, for $0 < \alpha < 1$.

Since the distribution of $T_{r,s}$ is a discrete distribution, we cannot attain the same size for different parameter values with a non-randomized test. To perform a more meaningful comparison of the power of the tests, a randomized testing procedure is adapted that is described below.

Let $\Phi$ be a randomized test given by,

$$\Phi = \begin{cases} 1, & if\ T_{r,s} \geq c, \\ \frac{\alpha - \alpha_1}{\alpha_2 - \alpha_1}, & if\ T_{r,s} = (c-1), \\ 0, & othwerwise, \end{cases} \tag{7}$$

Where $\alpha_1 = P[T_{r,s} \geq c]$, $\alpha_2 = P[T_{r,s} \geq (c-1)]$, $0 < \alpha_1 \leq \alpha \leq \alpha_2 < 1$. Then the size of the test is obtained as $\sup E(\Phi|H_0) \leq \alpha$. Limiting distribution of $T_{r,s}$ under $H_0$ is presented in Theorem 3.2.

**Theorem 3.2.** For $0 \leq f_{pi}, f_{ei} \leq m$, $N_1 = \sum_1^r f_{pi}$, $N_2 = \sum_1^s f_{ei}$, and $N = N_1 + N_2$, and given $(r, s)$, the null distribution of $T_{r,s} = P_r + E_s$ is given by

$F_{r,s}(t|H_0)$

$= \sum_{k=0}^{t} \sum_{i=0}^{k} \sum_{N=0}^{\min(m, ri+s(k-i))} \sum_{N_1=0}^{\min(N,ri)} I((N-N_1) \leq s(k-i)) w(N_1, r, i, 0)\ w((N-N_1), s, (k-i), 0) \frac{\binom{m-N+n-r-s}{n-r-s}}{\binom{m+n}{n}}$, for $0 \leq t \leq m$,

$= 0$, otherwise,

where $I(.)$ is the standard indicator function, and $w(a, b, m_1, m_2)$ is the number of ways that '$a$' indistinguishable objects can be distributed among '$b$' distinguishable boxes with $m_1$ maximum and $m_2$ minimum objects in each box.

*Proof.* The proof is presented in Appendix A3.

As evident from Theorem 3.2, the null distribution of $T_{r,s}$ only depends on $\{m, n, r, s\}$ irrespective of $F$ and $G$, and therefore the proposed test is distribution-free under $H_0$. Computation of the exact distribution of $T_{r,s}$ is time consuming, particularly under the Lehmann alternative. Therefore, the critical values $c$ at $\alpha = 0.05$ are obtained for $r = s$ and $r \neq s$ with 100,000 Monte-Carlo simulations. These values are reported in Table 3-5. Note that $F_{r,s}(t|H_0) = F_{s,r}(t|H_0)$. Therefore, for $r \neq s$, critical values $c$ with $(\alpha_1, \alpha_2)$ are same when $r$ and $s$ values are interchanged.

**Table 3. Critical values $c$ and corresponding $\alpha_1$, $\alpha_2$ for $r = s$ at 5% level of significance.**

| $\rho$ | $m$ | $n$ | $r$ | $s$ | $c$ | $\alpha_1$ | $\alpha_2$ | $\rho$ | $m$ | $n$ | $r$ | $s$ | $c$ | $\alpha_1$ | $\alpha_2$ |
|---|---|---|---|---|---|---|---|---|---|---|---|---|---|---|---|
| 0.05 | 10 | 10 | 1 | 1 | 6 | 0.03 | 0.07 | 0.25 | 10 | 10 | 3 | 3 | 8 | 0.02 | 0.06 |
| 0.05 | 10 | 20 | 2 | 2 | 5 | 0.03 | 0.09 | 0.25 | 10 | 20 | 6 | 6 | 6 | 0.05 | 0.17 |
| 0.05 | 10 | 30 | 2 | 2 | 4 | 0.03 | 0.11 | 0.25 | 10 | 30 | 8 | 8 | 6 | 0.02 | 0.07 |
| 0.05 | 20 | 10 | 1 | 1 | 10 | 0.04 | 0.06 | 0.25 | 20 | 10 | 3 | 3 | 13 | 0.04 | 0.07 |
| 0.05 | 20 | 20 | 2 | 2 | 8 | 0.03 | 0.06 | 0.25 | 20 | 20 | 6 | 6 | 10 | 0.04 | 0.08 |
| 0.05 | 20 | 30 | 2 | 2 | 6 | 0.04 | 0.08 | 0.25 | 20 | 30 | 8 | 8 | 9 | 0.02 | 0.05 |
| 0.05 | 30 | 10 | 1 | 1 | 14 | 0.04 | 0.06 | 0.25 | 30 | 10 | 3 | 3 | 18 | 0.04 | 0.07 |
| 0.05 | 30 | 20 | 2 | 2 | 11 | 0.03 | 0.06 | 0.25 | 30 | 20 | 6 | 6 | 14 | 0.03 | 0.06 |
| 0.05 | 30 | 30 | 2 | 2 | 8 | 0.04 | 0.07 | 0.25 | 30 | 30 | 8 | 8 | 11 | 0.04 | 0.09 |
| 0.1 | 10 | 10 | 2 | 2 | 7 | 0.03 | 0.09 | 0.35 | 10 | 10 | 4 | 4 | 8 | 0.03 | 0.10 |
| 0.1 | 10 | 20 | 3 | 3 | 6 | 0.01 | 0.05 | 0.35 | 10 | 20 | 8 | 8 | 7 | 0.02 | 0.07 |
| 0.1 | 10 | 30 | 4 | 4 | 5 | 0.02 | 0.09 | 0.35 | 10 | 30 | 11 | 11 | 6 | 0.02 | 0.11 |
| 0.1 | 20 | 10 | 2 | 2 | 12 | 0.04 | 0.07 | 0.35 | 20 | 10 | 4 | 4 | 14 | 0.03 | 0.06 |
| 0.1 | 20 | 20 | 3 | 3 | 9 | 0.02 | 0.06 | 0.35 | 20 | 20 | 8 | 8 | 11 | 0.02 | 0.06 |
| 0.1 | 20 | 30 | 4 | 4 | 7 | 0.04 | 0.10 | 0.35 | 20 | 30 | 11 | 11 | 9 | 0.03 | 0.09 |
| 0.1 | 30 | 10 | 2 | 2 | 17 | 0.04 | 0.06 | 0.35 | 30 | 10 | 4 | 4 | 19 | 0.05 | 0.08 |
| 0.1 | 30 | 20 | 3 | 3 | 12 | 0.03 | 0.05 | 0.35 | 30 | 20 | 8 | 8 | 15 | 0.02 | 0.05 |
| 0.1 | 30 | 30 | 4 | 4 | 10 | 0.03 | 0.06 | 0.35 | 30 | 30 | 11 | 11 | 12 | 0.03 | 0.07 |

**Table 4. Critical values $c$ and corresponding $\alpha_1$, $\alpha_2$ for $r \neq s$ and $m = n = 10$ at 5% level of significance.**

| $m = n = 10$ | | | | | | | | | | | | |
|---|---|---|---|---|---|---|---|---|---|---|---|---|
| | $s = 1$ | | | 2 | | | 3 | | | 4 | | |
| $r$ | $c$ | $\alpha_1$ | $\alpha_2$ | $c$ | $\alpha_1$ | $\alpha_2$ | $c$ | $\alpha_1$ | $\alpha_2$ | $c$ | $\alpha_1$ | $\alpha_2$ |
| 1 | 6 | 0.03 | 0.07 | 7 | 0.02 | 0.05 | 7 | 0.02 | 0.07 | 7 | 0.03 | 0.09 |
| 2 | 7 | 0.02 | 0.05 | 7 | 0.03 | 0.09 | 7 | 0.04 | 0.12 | 8 | 0.02 | 0.06 |
| 3 | 7 | 0.03 | 0.07 | 7 | 0.04 | 0.12 | 8 | 0.02 | 0.06 | 8 | 0.02 | 0.08 |
| 4 | 7 | 0.03 | 0.09 | 8 | 0.02 | 0.06 | 8 | 0.02 | 0.08 | 8 | 0.03 | 0.10 |

**Table 5. Critical values $c$ and corresponding $\alpha_1$, $\alpha_2$ for $r \neq s$ and $m = n = 20$ at 5% level of significance.**

| $m = n = 20$ | | | | | | | | | | | | |
|---|---|---|---|---|---|---|---|---|---|---|---|---|
| | $s = 1$ | | | 2 | | | 3 | | | 4 | | |
| $r$ | $c$ | $\alpha_1$ | $\alpha_2$ | $c$ | $\alpha_1$ | $\alpha_2$ | $c$ | $\alpha_1$ | $\alpha_2$ | $c$ | $\alpha_1$ | $\alpha_2$ |
| 1 | 6 | 0.04 | 0.09 | 7 | 0.04 | 0.08 | 8 | 0.02 | 0.05 | 8 | 0.03 | 0.07 |
| 2 | 7 | 0.04 | 0.08 | 8 | 0.03 | 0.06 | 8 | 0.04 | 0.09 | 9 | 0.02 | 0.05 |
| 3 | 8 | 0.02 | 0.05 | 8 | 0.04 | 0.09 | 9 | 0.03 | 0.06 | 9 | 0.03 | 0.07 |
| 4 | 8 | 0.03 | 0.07 | 9 | 0.02 | 0.05 | 9 | 0.03 | 0.07 | 9 | 0.04 | 0.09 |
| 5 | 8 | 0.04 | 0.08 | 9 | 0.03 | 0.06 | 9 | 0.04 | 0.09 | 10 | 0.02 | 0.05 |
| 6 | 8 | 0.04 | 0.09 | 9 | 0.03 | 0.07 | 9 | 0.05 | 0.10 | 10 | 0.03 | 0.06 |
| 7 | 8 | 0.05 | 0.10 | 9 | 0.04 | 0.09 | 10 | 0.02 | 0.05 | 10 | 0.03 | 0.07 |
| 8 | 9 | 0.02 | 0.06 | 9 | 0.04 | 0.09 | 10 | 0.03 | 0.06 | 10 | 0.03 | 0.08 |
| | $s = 5$ | | | 6 | | | 7 | | | 8 | | |
| 1 | 8 | 0.04 | 0.08 | 8 | 0.04 | 0.09 | 8 | 0.05 | 0.10 | 9 | 0.02 | 0.06 |
| 2 | 9 | 0.03 | 0.06 | 9 | 0.03 | 0.07 | 9 | 0.04 | 0.09 | 9 | 0.04 | 0.09 |
| 3 | 9 | 0.04 | 0.09 | 9 | 0.05 | 0.10 | 10 | 0.02 | 0.05 | 10 | 0.03 | 0.06 |
| 4 | 10 | 0.02 | 0.05 | 10 | 0.03 | 0.06 | 10 | 0.03 | 0.07 | 10 | 0.03 | 0.08 |
| 5 | 10 | 0.03 | 0.06 | 10 | 0.03 | 0.07 | 10 | 0.04 | 0.08 | 10 | 0.04 | 0.09 |
| 6 | 10 | 0.03 | 0.07 | 10 | 0.04 | 0.08 | 10 | 0.04 | 0.09 | 10 | 0.05 | 0.11 |
| 7 | 10 | 0.04 | 0.08 | 10 | 0.04 | 0.09 | 10 | 0.05 | 0.11 | 11 | 0.02 | 0.05 |
| 8 | 10 | 0.04 | 0.09 | 10 | 0.05 | 0.11 | 11 | 0.02 | 0.05 | 11 | 0.02 | 0.06 |

The limiting distribution of $T_{r,s}$ under $H_0$ is provided in Theorem 3.3.

**Theorem 3.3.** For $m, n \to \infty$ with $\frac{m}{n} \to 1$, $1 < r, s < \infty$, and $0 \leq \sum_1^s f_{ei} + \sum_1^r f_{pi} = N < \infty$ for some given $N$, the asymptotic null distribution of $T_{r,s}$ is given by

$$P[T_{r,s} \leq t \mid H_0] \approx \sum_{k=0}^{t} \sum_{i=0}^{k} \sum_{N=0}^{\min(m, ri+s(k-i))} \sum_{N_1=0}^{\min(N, ri)} I((N - N_1) \leq s(k - i)) w(N_1, r, i, 0)\, w(N_2, s, (k - i), 0) \left(\frac{1}{2}\right)^{N+r+s}, \quad 0 \leq t \leq a,$$

$= 0$, otherwise,

*Proof.* The proof is presented in Appendix A3.

## 4. Distribution under Lehmann alternative

We derive the distribution of $T_{r,s}$ under Lehmann alternative, $H_1: G = F^\gamma$ for $\gamma > 0$. When $\gamma < 1$, the alternative hypothesis is given by $H_1: F < G$. This means that the *test* group $Y$ is stochastically smaller than the *training* group $X$. For $\gamma > 1$, we have $H_1: F > G$, i.e., the *test* group $Y$ is stochastically larger than the *training* group $X$.

**Theorem 4.1.** For $0 \leq f_{pi}, f_{ei} \leq m$, the joint probability mass function of $\{f_{p1}, f_{p2}, \ldots, f_{pr}, f_{e1}, f_{e2}, \ldots, f_{es}\}$ under $H_1$ is given by

$P[f_{p1}, f_{p2}, \ldots, f_{pr}, f_{e1}, f_{e2}, \ldots, f_{es} \mid H_1]$

$= K\gamma^{r+s} \sum_{l=0}^{n-r-s} \binom{n-r-s}{l} B(f_{p1} + \gamma, f_{p2} + 1)\, B(f_{p1} + f_{p2} + 2\gamma, f_{p3} + 1) \ldots B(\sum_{i=1}^{r-1} f_{pi} + (r-1)\gamma, f_{pr} + 1)\quad B(\sum_{i=1}^{r} f_{pi} + r\gamma + \gamma l,\ m - \sum_{i=1}^{r} f_{pi} - \sum_{i=1}^{s} f_{ei} + 1)\quad B(m + \gamma(n - s) - \sum_1^s f_{ei} + \gamma, f_{e1} + 1)\, B(m + \gamma(n - s) - \sum_2^s f_{ei} + 2\gamma, f_{e1} + 1) \ldots$

$B(m + \gamma(n - s) - \sum_{s-1}^{s} f_{ei} + (s - 1)\gamma, f_{e(s-1)} + 1)\, B(m + \gamma(n - s) - f_{es} + s\gamma, f_{es} + 1).$

$= 0$, otherwise.

where $0 \leq \sum_1^s f_{ei} + \sum_1^r f_{pi} = N \leq m$, $0 < r + s \leq n$, and

$$K = \frac{m!\, n!}{\prod_1^r f_{pi}!\, (m-\sum_1^r f_{pi}-\sum_1^s f_{ei})!\, \prod_1^s f_{ei}!\, (n-r-s)!}.$$

*Proof.* The proof is presented in Appendix A1.

The distribution of $T_{r,s}$ under Lehmann alternative can be obtained as

$$P[T_{r,s} \leq t \,|H_1] = \sum_{k=0}^{t} \sum_{i=0}^{k} P[P_r = i, E_s = k - i \mid H_1], \, 0 \leq t \leq m, \tag{8}$$

where,

$P[P_r = i, E_s = k - i \mid H_1]$

$$= \sum_{\substack{f_{p1},f_{p2},\ldots,f_{pr},\ldots,f_{e1},f_{e2},\ldots,f_{es} \\ P_r=i\,;\, E_s=k-i}} P[f_{p1}, f_{p2}, \ldots, f_{pr},\, f_{e1}, f_{e2}, \ldots, f_{es} \mid H_1]$$

Note that, the joint probability mass function of $\{f_{p1}, f_{p2}, \ldots, f_{pr},\, f_{e1}, f_{e2}, \ldots, f_{es}\}$ under $H_1$ depends on $\{m, n, r, s, \gamma\}$. This implies that the proposed test is distribution-free under Lehmann alternative.

### 4.1. Power comparison under Lehmann alternative

Power of the proposed test is obtained by the randomized test procedure described in (7). With the randomized test $\Phi$, the power of a test is given by $\inf E(\Phi|H_1)$. The proposed test is said to be finite sample unbiased if $\inf E(\Phi|H_1) \geq \alpha$ for finite $\{m, n\}$. From (7), we can write,

$$E(\Phi|H_1) = P[T_{r,s} \geq c|H_1] + \frac{\alpha-\alpha_1}{\alpha_2-\alpha_1} P[T_{r,s} = c - 1|H_1] \tag{9}$$

For $m = n = 10, 20, 30$, power of the $T_{r,s}$ test is obtained under Lehmann alternative for each of $\gamma = \{1/10, 1/9, 1/8, …,8, 9, 10\}$. The results are presented in Table 6. For unequal sample sizes, power values are presented in Table 7. For all power comparisons, 100,000 Monte-Carlo simulations are performed.

From Table 6 and 7, it can be observed that the power values increase for larger deviations in $\gamma$ from 1. To test if the training group is stochastically larger than the test group, i.e., against $\gamma < 1$, higher power is attained for $r > 1$. For example, with $m = n = 30$ and $\gamma \leq 1/6$ in Table 6, the most powerful test is for $r = s = 4$. Power curves of the $T_{r,s}$ test for $m = n = 30$ are illustrated in Fig. 1. As it can be observed from Fig. 1, $r = s > 1$ yield more power when testing for the training group being stochastically larger than the test group.

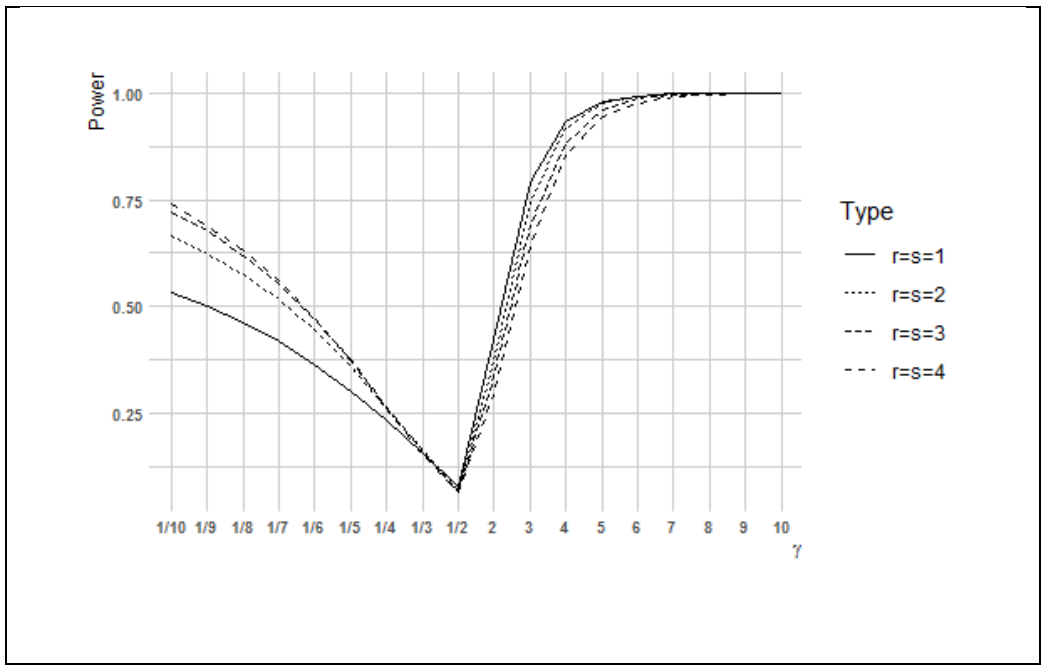

**Fig. 1. Power curves of $T_{r,s}$ tests for $m = n = 30$ under Lehman alternative with $\gamma > 1$ and $\gamma < 1$.**

**Table 6. Power of the $T_{r,s}$ test at 5% level of significance for $m = n = 10, 20, 30$ and $r = s$.**

| $m = n$ | 10 | | | | 20 | | | | 30 | | | |
|---|---|---|---|---|---|---|---|---|---|---|---|---|
| $r = s$ | 1 | 2 | 3 | 4 | 1 | 2 | 3 | 4 | 1 | 2 | 3 | 4 |
| $c$ | 6 | 7 | 8 | 8 | 6 | 8 | 9 | 9 | 7 | 8 | 9 | 10 |
| $\gamma$ | | | | | | | | | | | | |
| 1/10 | 0.501 | 0.517 | 0.448 | 0.380 | 0.524 | 0.642 | 0.662 | 0.630 | 0.532 | 0.666 | 0.720 | 0.740 |
| 1/9 | 0.463 | 0.480 | 0.412 | 0.346 | 0.492 | 0.597 | 0.612 | 0.585 | 0.500 | 0.624 | 0.678 | 0.688 |
| 1/8 | 0.424 | 0.434 | 0.371 | 0.307 | 0.453 | 0.548 | 0.558 | 0.528 | 0.460 | 0.577 | 0.620 | 0.632 |
| 1/7 | 0.379 | 0.384 | 0.323 | 0.266 | 0.408 | 0.490 | 0.495 | 0.467 | 0.417 | 0.517 | 0.552 | 0.558 |
| 1/6 | 0.333 | 0.327 | 0.215 | 0.222 | 0.357 | 0.421 | 0.420 | 0.388 | 0.362 | 0.444 | 0.471 | 0.475 |
| 1/5 | 0.270 | 0.264 | 0.159 | 0.176 | 0.296 | 0.339 | 0.332 | 0.306 | 0.302 | 0.360 | 0.374 | 0.373 |
| 1/4 | 0.205 | 0.192 | 0.104 | 0.134 | 0.225 | 0.247 | 0.237 | 0.214 | 0.232 | 0.262 | 0.265 | 0.261 |
| 1/3 | 0.136 | 0.122 | 0.063 | 0.094 | 0.148 | 0.152 | 0.140 | 0.128 | 0.154 | 0.162 | 0.157 | 0.150 |
| 1/2 | 0.070 | 0.066 | 0.108 | 0.064 | 0.073 | 0.068 | 0.064 | 0.062 | 0.075 | 0.069 | 0.065 | 0.064 |
| 2 | 0.189 | 0.138 | 0.210 | 0.086 | 0.324 | 0.266 | 0.225 | 0.189 | 0.424 | 0.369 | 0.330 | 0.292 |
| 3 | 0.385 | 0.269 | 0.325 | 0.171 | 0.659 | 0.566 | 0.487 | 0.433 | 0.793 | 0.744 | 0.689 | 0.643 |
| 4 | 0.551 | 0.403 | 0.428 | 0.269 | 0.844 | 0.774 | 0.700 | 0.648 | 0.934 | 0.917 | 0.883 | 0.855 |
| 5 | 0.678 | 0.520 | 0.521 | 0.368 | 0.932 | 0.888 | 0.835 | 0.792 | 0.980 | 0.975 | 0.959 | 0.946 |
| 6 | 0.770 | 0.614 | 0.601 | 0.457 | 0.969 | 0.944 | 0.911 | 0.881 | 0.993 | 0.993 | 0.987 | 0.978 |
| 7 | 0.832 | 0.690 | 0.601 | 0.531 | 0.986 | 0.971 | 0.950 | 0.931 | 0.998 | 0.998 | 0.995 | 0.992 |
| 8 | 0.876 | 0.753 | 0.664 | 0.596 | 0.993 | 0.985 | 0.972 | 0.960 | 0.999 | 0.999 | 0.998 | 0.997 |
| 9 | 0.908 | 0.797 | 0.714 | 0.648 | 0.997 | 0.992 | 0.984 | 0.976 | 0.999 | 0.999 | 0.999 | 0.999 |
| 10 | 0.929 | 0.834 | 0.758 | 0.694 | 0.998 | 0.996 | 0.991 | 0.985 | 0.999 | 0.999 | 0.999 | 0.999 |

**Table 7. Power of the $T_{r,s}$-test at 5% level of significance for $m = 30$, $n = 10, 25$ and $\gamma = 5$, 1/5.**

|  | $\gamma = 5$ | | $\gamma = 1/5$ | |
|---|---|---|---|---|
| $r = s$ | $n = 10$ | $n = 25$ | $n = 10$ | $n = 25$ |
| 1 | 0.845 | 0.959 | 0.331 | 0.307 |
| 2 | 0.714 | 0.937 | 0.344 | 0.365 |
| 3 | 0.632 | 0.894 | 0.303 | 0.377 |
| 4 | 0.574 | 0.865 | 0.270 | 0.358 |

### 4.2. Power comparison with exceedance-precedence tests:

Balakrishnan and Frattina (2000) defined the maximal precedence test statistic as a generalization of the traditional precedence test. It is defined as the maximum frequency of $X$-failures occurring between every consecutive ordered $Y$-failure, till the $r^{th}$ $Y$ failure, i.e.,

$$Q_r = \max\{f_{p1}, f_{p2}, f_{p3}, \ldots, f_{pr}\}.$$

Stoimenova and Balakrishnan (2017) proposed a Šidák-type test as the summation of precedence and exceedance statistics. Suppose,

$A_s$ = the number of $Y$-observations exceeding $X_{(m-s+1)}$,

$B_r$ = the number of $X$-observations preceding $Y_{(r)}$,

where $1 \leq r \leq n$ and $1 \leq s \leq m$.

Then the Šidák-type precedence-exceedance statistic by Stoimenova and Balakrishnan (2017) is defined as

$V_r = A_r + B_r$, for $r = s, m = n$.

For simplicity, let us denote the $T_{r,s}$ test as $T_r$ test for $r = s$. In Table 8, power comparison of $T_r$ test, $V_r$ test, and $Q_r$ test at 5% level of significance is provided under Lehmann alternative. Sample size is taken as $m = n = 25$. It can be observed that the $V_r$ test and $Q_r$ test would not reject $H_0$ at 5% level of significance when testing for $\gamma < 1$. However, $T_r$ test truly rejects $H_0$ at 5% level of significance for both $\gamma > 1$ and $\gamma < 1$. In fact, power of the $T_r$ test increases for $r > 1$ when testing for $\gamma < 1$, i.e. the *test* group is stochastically smaller than the *training* group.

This is important because we may not have information about $\gamma$ in real life. If the *test* group is stochastically smaller than the *training* group, and we perform $V_r$ test or $Q_r$ test, the test statistics would fail to reject $H_0$ at 5% level of significance. In contrast, $T_r$ test is more likely to truly reject $H_0$ for $r = 3$ or 4 for sample size 20 or higher.

To compare power values of $T_r$ test, $V_r$ test, and $Q_r$ test under some popular lifetime distributions, we consider exponential distribution with rate parameter $\lambda = 1$, Weibull distributions with scale parameter $\eta = 1$ and shape parameter $\nu = 0.5, 2.5$. Results are reported in Table 9. It can be observed that the $T_r$ test is highly powerful to reject $H_0$ at 5% level of significance for both increase and decrease in $\lambda$ for exponential distribution and $\eta$ for Weibull distribution. On the contrary, $V_r$ test and $Q_r$ test fail to reject $H_0$ when $\lambda > 1$ and $\eta < 1$.

Power comparison of $T_r$ test with $V_r$ and $Q_r$ tests at 5% level of significance under Exponential(1) and Weibull(2.5,1) distribution is illustrated in Fig. 2 and 3, respectively. It can be observed from Fig. 3 that $T_r$ test is more powerful than $Q_r$ test at 5% level of significance for all $r = s$ and for all shift in scale parameter $\eta$ of Weibull(2.5,1) distribution.

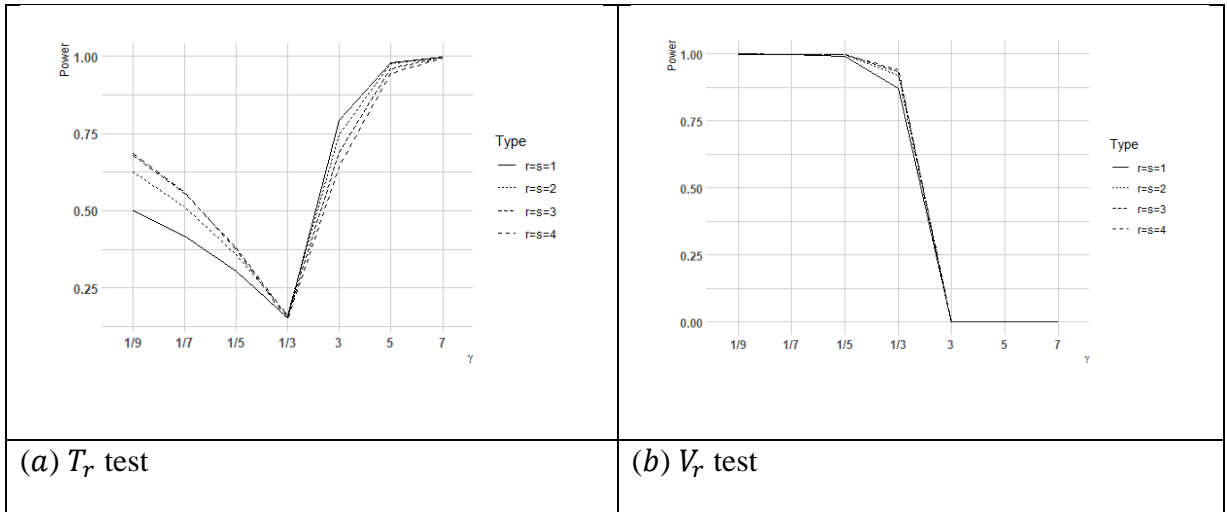

| (a) $T_r$ test | (b) $V_r$ test |

**Fig. 2.** Power comparison of $T_r$ and $V_r$ test at 5% level of significance under Exponential(1) for $m = n = 30$.

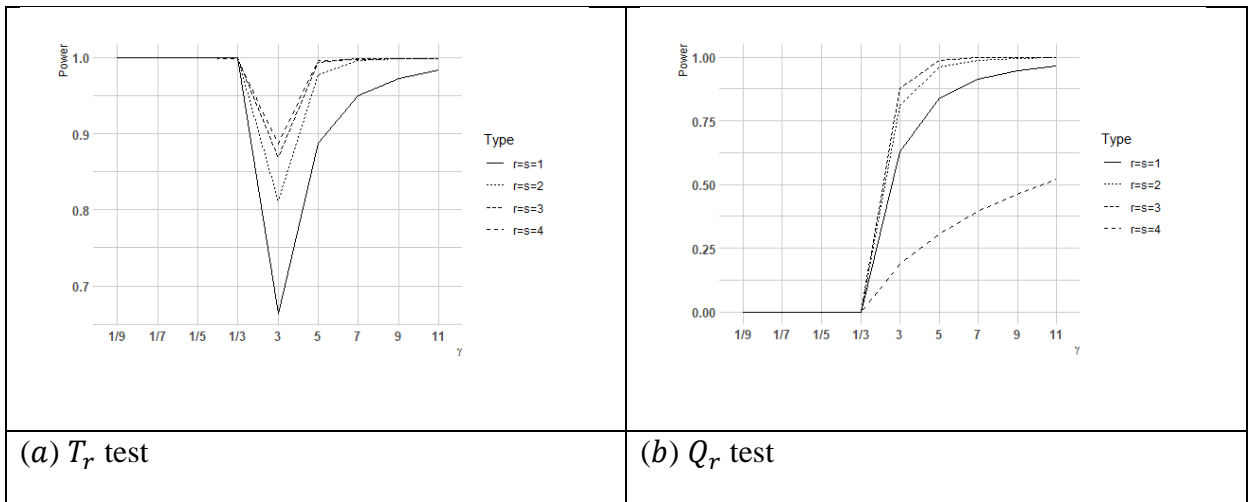

| (a) $T_r$ test | (b) $Q_r$ test |

**Fig. 3.** Power comparison of $T_r$ and $Q_r$ test at 5% level of significance under Weibull(2.5,1) for $m = n = 30$.

**Table 8. Power comparison of $T_r$ test, $V_r$ test, and $Q_r$ test at 5% level of significance for $m = n = 25$ under Lehmann alternative.**

|       | $T_r$ |       |       |       | $V_r$ |       |       |       | $Q_r$ |       |       |       |
|-------|-------|-------|-------|-------|-------|-------|-------|-------|-------|-------|-------|-------|
| $r=s$ | 1     | 2     | 3     | 4     | 1     | 2     | 3     | 4     | 1     | 2     | 3     | 4     |
| $c$   | 6     | 8     | 9     | 10    | 7     | 10    | 13    | 16    | 5     | 5     | 6     | 6     |
| $\gamma$ |    |       |       |       |       |       |       |       |       |       |       |       |
| 1/10  | 0.535 | 0.657 | 0.700 | 0.700 | 0.000 | 0.000 | 0.000 | 0.000 | 0.000 | 0.000 | 0.000 | 0.000 |
| 1/9   | 0.495 | 0.613 | 0.654 | 0.653 | 0.000 | 0.000 | 0.000 | 0.000 | 0.000 | 0.000 | 0.000 | 0.000 |
| 1/8   | 0.458 | 0.563 | 0.600 | 0.594 | 0.000 | 0.000 | 0.000 | 0.000 | 0.000 | 0.000 | 0.000 | 0.000 |
| 1/7   | 0.415 | 0.504 | 0.529 | 0.526 | 0.000 | 0.000 | 0.000 | 0.000 | 0.000 | 0.000 | 0.000 | 0.000 |
| 1/6   | 0.360 | 0.434 | 0.451 | 0.443 | 0.000 | 0.000 | 0.000 | 0.000 | 0.000 | 0.000 | 0.000 | 0.000 |
| 1/5   | 0.300 | 0.351 | 0.357 | 0.348 | 0.000 | 0.000 | 0.000 | 0.000 | 0.000 | 0.000 | 0.000 | 0.000 |
| 1/4   | 0.231 | 0.258 | 0.252 | 0.242 | 0.000 | 0.000 | 0.000 | 0.000 | 0.000 | 0.000 | 0.000 | 0.000 |
| 1/3   | 0.154 | 0.159 | 0.149 | 0.144 | 0.000 | 0.000 | 0.000 | 0.000 | 0.000 | 0.000 | 0.000 | 0.000 |
| 1/2   | 0.074 | 0.069 | 0.064 | 0.064 | 0.001 | 0.000 | 0.000 | 0.000 | 0.000 | 0.000 | 0.000 | 0.000 |
| 2     | 0.379 | 0.325 | 0.274 | 0.246 | 0.506 | 0.564 | 0.592 | 0.603 | 0.542 | 0.549 | 0.506 | 0.486 |
| 3     | 0.734 | 0.666 | 0.599 | 0.552 | 0.838 | 0.884 | 0.899 | 0.904 | 0.862 | 0.879 | 0.846 | 0.835 |
| 4     | 0.903 | 0.865 | 0.810 | 0.778 | 0.951 | 0.973 | 0.976 | 0.978 | 0.961 | 0.972 | 0.960 | 0.956 |
| 5     | 0.965 | 0.947 | 0.918 | 0.895 | 0.985 | 0.993 | 0.994 | 0.994 | 0.989 | 0.993 | 0.989 | 0.989 |
| 6     | 0.986 | 0.979 | 0.964 | 0.952 | 0.995 | 0.998 | 0.998 | 0.998 | 0.997 | 0.999 | 0.997 | 0.997 |
| 7     | 0.995 | 0.991 | 0.983 | 0.978 | 0.998 | 0.999 | 0.999 | 0.999 | 0.999 | 0.999 | 0.999 | 0.999 |

**Table 9. Power comparison of $T_r$ test, $V_r$ test, and $Q_r$ test at 5% level of significance for $m = n = 30$ under different lifetime distributions.**

|  |  | $T_r$ | | | | $V_r$ | | | | $Q_r$ | | | |
|---|---|---|---|---|---|---|---|---|---|---|---|---|---|
| $r = s$ | | 1 | 2 | 3 | 4 | 1 | 2 | 3 | 4 | 1 | 2 | 3 | 4 |
| $c$ | | 7 | 8 | 9 | 10 | 7 | 10 | 13 | 16 | 5 | 6 | 6 | 6 |
| Exp(1) | $\lambda$ | | | | | | | | | | | | |
| | 1/9 | 0.499 | 0.625 | 0.679 | 0.685 | 0.999 | 1.000 | 1.000 | 1.000 | 0.623 | 0.805 | 0.880 | 0.924 |
| | 1/7 | 0.418 | 0.513 | 0.555 | 0.558 | 0.999 | 0.999 | 0.999 | 0.999 | 0.555 | 0.728 | 0.810 | 0.865 |
| | 1/5 | 0.303 | 0.358 | 0.379 | 0.373 | 0.992 | 0.997 | 0.998 | 0.998 | 0.446 | 0.602 | 0.680 | 0.739 |
| | 1/3 | 0.154 | 0.162 | 0.157 | 0.148 | 0.872 | 0.920 | 0.935 | 0.941 | 0.280 | 0.376 | 0.424 | 0.466 |
| | 3 | 0.792 | 0.745 | 0.689 | 0.641 | 0.000 | 0.000 | 0.000 | 0.000 | 0.002 | 0.002 | 0.001 | 0.001 |
| | 5 | 0.980 | 0.975 | 0.959 | 0.944 | 0.000 | 0.000 | 0.000 | 0.000 | 0.000 | 0.000 | 0.000 | 0.000 |
| | 7 | 0.998 | 0.998 | 0.996 | 0.993 | 0.000 | 0.000 | 0.000 | 0.000 | 0.000 | 0.000 | 0.000 | 0.000 |
| Weibull(0.5,1) | $\eta$ | | | | | | | | | | | | |
| | 1/9 | 0.786 | 0.747 | 0.693 | 0.645 | 0.000 | 0.000 | 0.000 | 0.000 | 0.003 | 0.002 | 0.001 | 0.001 |
| | 1/7 | 0.685 | 0.638 | 0.582 | 0.530 | 0.000 | 0.000 | 0.000 | 0.000 | 0.004 | 0.002 | 0.002 | 0.002 |
| | 1/5 | 0.528 | 0.476 | 0.426 | 0.380 | 0.000 | 0.000 | 0.000 | 0.000 | 0.006 | 0.004 | 0.004 | 0.003 |
| | 1/3 | 0.290 | 0.256 | 0.228 | 0.201 | 0.002 | 0.000 | 0.000 | 0.000 | 0.014 | 0.011 | 0.009 | 0.008 |
| | 3 | 0.056 | 0.051 | 0.051 | 0.052 | 0.403 | 0.455 | 0.486 | 0.496 | 0.138 | 0.163 | 0.183 | 0.189 |
| | 5 | 0.090 | 0.087 | 0.085 | 0.082 | 0.656 | 0.726 | 0.755 | 0.767 | 0.199 | 0.249 | 0.282 | 0.304 |
| | 7 | 0.123 | 0.124 | 0.123 | 0.114 | 0.798 | 0.857 | 0.879 | 0.885 | 0.246 | 0.315 | 0.361 | 0.394 |
| | 9 | 0.152 | 0.160 | 0.158 | 0.151 | 0.873 | 0.922 | 0.937 | 0.940 | 0.283 | 0.372 | 0.426 | 0.463 |
| | 11 | 0.179 | 0.193 | 0.196 | 0.187 | 0.919 | 0.954 | 0.964 | 0.966 | 0.316 | 0.415 | 0.478 | 0.519 |
| Weibull(2.5,1) | $\eta$ | | | | | | | | | | | | |
| | 1/9 | 1.000 | 1.000 | 1.000 | 1.000 | 0.000 | 0.000 | 0.000 | 0.000 | 0.000 | 0.000 | 0.000 | 0.000 |
| | 1/7 | 1.000 | 1.000 | 1.000 | 1.000 | 0.000 | 0.000 | 0.000 | 0.000 | 0.000 | 0.000 | 0.000 | 0.001 |
| | 1/5 | 1.000 | 1.000 | 1.000 | 1.000 | 0.000 | 0.000 | 0.000 | 0.000 | 0.000 | 0.000 | 0.000 | 0.003 |
| | 1/3 | 1.000 | 1.000 | 1.000 | 0.999 | 0.000 | 0.000 | 0.000 | 0.000 | 0.000 | 0.000 | 0.000 | 0.008 |
| | 3 | 0.664 | 0.812 | 0.869 | 0.886 | 1.000 | 1.000 | 1.000 | 1.000 | 0.628 | 0.808 | 0.878 | 0.189 |
| | 5 | 0.888 | 0.977 | 0.993 | 0.996 | 1.000 | 1.000 | 1.000 | 1.000 | 0.839 | 0.963 | 0.988 | 0.304 |
| | 7 | 0.950 | 0.996 | 0.999 | 0.997 | 1.000 | 1.000 | 1.000 | 1.000 | 0.914 | 0.989 | 0.998 | 0.394 |
| | 9 | 0.973 | 0.999 | 0.999 | 0.999 | 1.000 | 1.000 | 1.000 | 1.000 | 0.947 | 0.995 | 0.999 | 0.463 |
| | 11 | 0.984 | 0.999 | 0.999 | 0.999 | 1.000 | 1.000 | 1.000 | 1.000 | 0.964 | 0.998 | 0.999 | 0.519 |

**Example 4.1**. Let us consider the same dataset on failure voltages for Type I and Type II insulation presented in Example 3.1. Let us first consider the Type I sample as *training* sample and Type II sample as *test* sample. We perform $T_r$ test and $V_r$ test at 5% level of significance against $H_1: G = F^\gamma, \gamma > 1$. Test statistic values for $r = 1, 2, 3, 4$ with corresponding critical values are presented in Table 10. It can be observed from Table 10 that, $T_r$ test reject $H_0$ at 5% level of significance for $r = 3$ and 4, whereas $V_r$ test reject $H_0$ for all $r = 1, 2, 3, 4$.

**Table 10. Testing $H_0: G = F$ with different $T_r$ and $V_r$ tests.**

| Test | Test statistic | Critical value | Test | Test statistic | Critical value |
|------|----------------|----------------|------|----------------|----------------|
| $T_1$ | 3 | 6 | $V_1$ | 13 | 7 |
| $T_2$ | 3 | 8 | $V_2$ | 20 | 10 |
| $T_3$ | 10 | 9 | $V_3$ | 32 | 13 |
| $T_4$ | 10 | 9 | $V_4$ | 32 | 16 |

Suppose an experimenter considered Type II sample as *training* sample and Type I sample as *test* sample. Test statistic values for $r = 1, 2, 3, 4$ with corresponding critical values are presented in Table 11. It can be observed that $T_r$ test still rejects $H_0$ at 5% level of significance for all $r$, whereas $V_r$ tests fail to reject $H_0$ for any $r$ when *test* and *training* samples are interchanged. In this context, $T_r$ test is more appropriate than $Q_r$ or $V_r$ test because it is suitable for two-sided alternative.

**Table 11. Testing $H_0: G = F$ with different $T_r$ and $V_r$ tests.**

| Test | Test statistic | Critical value | Test | Test statistic | Critical value |
|------|----------------|----------------|------|----------------|----------------|
| $T_1$ | 10 | 6 | $V_1$ | 0 | 7 |
| $T_2$ | 10 | 8 | $V_2$ | 0 | 10 |
| $T_3$ | 10 | 9 | $V_3$ | 0 | 13 |
| $T_4$ | 11 | 9 | $V_4$ | 1 | 16 |

## 5. Conclusion

In this article, a Šidák-type test is developed based on maximal precedence and maximal exceedance statistics. We have obtained the joint distribution of maximal precedence and maximal exceedance statistics. Exact distributions of the proposed test statistic is obtained under the null and Lehmann alternative. The proposed test does not require any distributional assumption for the lifetime data which makes it distribution-free under the null as well as under the Lehmann alternative. With extensive numerical study and a real-life example, it is shown that the maximal precedence tests and Šidák-type precedence-exceedance tests are finite sample biased for a certain class of alternatives. The proposed test is free from such bias and applicable in case of a two-sided alternative. In such context, the proposed test is more suitable than contending precedence-exceedance type tests.

**Declaration of interest:** none.

**Appendix**

**A1. Proof of Theorem 4.1.**

We begin by proving Theorem 4.1. on the joint probability mass function of $\{f_{p1}, f_{p2}, \ldots, f_{pr}, f_{e1}, f_{e2}, \ldots, f_{es}\}$ under $H_1: G = F^\gamma$, $\gamma > 0$. Subsequent null distribution of $\{f_{p1}, f_{p2}, \ldots, f_{pr}, f_{e1}, f_{e2}, \ldots, f_{es}\}$ can be easily obtained by taking $\gamma = 1$ in the joint probability $P[f_{p1}, f_{p2}, \ldots, f_{pr}, f_{e1}, f_{e2}, \ldots, f_{es}]$.

We have $\{f_{p1}, f_{p2}, \ldots, f_{pr}, f_{e1}, f_{e2}, \ldots, f_{es}\}$ such that $0 \leq f_{pi}, f_{ei} \leq m$,

Under $H_1$,

$P_a = P[f_{p1}, f_{p2}, \ldots, f_{pr}, \ldots, f_{e1}, f_{e2}, \ldots, f_{es} | H_1]$

$$= \int_{\boldsymbol{y}} P[f_{p1}, f_{p2}, \ldots, f_{pr}, \ldots, f_{e1}, f_{e2}, \ldots, f_{es} | y_{(1)}, y_{(2)}, \ldots, y_{(r)}, y_{(n-s+1)}, \ldots, y_{(n)}] \, f_{\boldsymbol{y}} \, d\boldsymbol{y}$$

$$= K_1 \int F^{f_{p1}}(y_{(1)}) [F(y_{(2)}) - F(y_{(1)})]^{f_{p2}} [F(y_{(3)}) - F(y_{(2)})]^{f_{p3}} \ldots [F(y_{(r)}) - F(y_{(r-1)})]^{f_{pr}} [F(y_{(n-s+1)}) - F(y_{(r)})]^{m - \sum_1^r f_{pi} - \sum_1^s f_{ei}} [F(y_{(n-s+2)}) - F(y_{(n-s+1)})]^{f_{e1}} [F(y_{(n-s+3)}) - F(y_{(n-s+2)})]^{f_{e2}} \ldots [F(y_{(n)}) - F(y_{(n-1)})]^{f_{e(s-1)}} [1 - F(y_{(n)})]^{f_{es}} f_{\boldsymbol{y}} \, d\boldsymbol{y},$$

where,

$$\boldsymbol{y} = \{y_{(1)}, y_{(2)}, \ldots, y_{(r)}, y_{(n-s+1)}, \ldots, y_{(n)}\}, \quad K_1 = \frac{m!}{\prod_1^r f_{pi}! \, (m - \sum_1^r f_{pi} - \sum_1^s f_{ei})! \prod_1^s f_{ei}!},$$

and,

$$f_{\boldsymbol{y}} = \frac{n!}{(n-r-s)!} [G(y_{(n-s+1)}) - G(y_{(r)})]^{n-r-s} g(y_{(1)}) g(y_{(2)}) \ldots g(y_{(r)}) g(y_{(n-s+1)}) g(y_{(n-s+2)}) \ldots g(y_{(n)}).$$

Let us denote $K_2 = \frac{n!}{(n-r-s)!}$.

Taking the first derivative of $G(x) = F^{\gamma}(x)$, we get $g(x) = \gamma F^{\gamma - 1}(x) f(x)$. Then,

$$P[f_{p1}, f_{p2}, \ldots, f_{pr}, \ldots, f_{e1}, f_{e2}, \ldots, f_{es}] = K \int \left\{ \prod_{i=1}^r [F(y_{(i)}) - F(y_{(i-1)})]^{f_{pi}} \right\} [F(y_{(n-s+1)}) - F(y_{(r)})]^{m - \sum_{i=1}^r f_{pi} - \sum_{i=1}^s f_{ei}} \left\{ \prod_{i=1}^s [F(y_{(n-s+i+1)}) - F(y_{(n-s+i)})]^{f_{ei}} \right\} [F^{\gamma}(y_{(n-s+1)}) - F^{\gamma}(y_{(r)})]^{n-r-s} \left\{ \prod_{i=1}^r \gamma F^{\gamma - 1}(y_{(i)}) f(y_{(i)}) \right\} \left\{ \prod_{i=1}^s \gamma F^{\gamma - 1}(y_{(n-s+i)}) f(y_{(n-s+i)}) \right\} d\boldsymbol{y},$$

where $F(y_0) = 0$ and $F(y_{(n+1)}) = 1$ and $K = K_1 K_2 = \frac{m! \, n!}{\prod_1^r f_{pi}! \, (m - \sum_1^r f_{pi} - \sum_1^s f_{ei})! \prod_1^s f_{ei}! \, (n-r-s)!}$.

With probability integral transformation, we can write

$$P[f_{p1}, f_{p2}, \ldots, f_{pr}, \ldots, f_{e1}, f_{e2}, \ldots, f_{es}] = K \int \left\{ \prod_{i=1}^{r} [u_{(i)} - u_{(i-1)}]^{f_{pi}} \right\} [u_{(n-s+1)} - u_{(r)}]^{m-\sum_{i=1}^{r} f_{pi} - \sum_{i=1}^{s} f_{ei}} \left\{ \prod_{i=1}^{s} [u_{(n-s+i+1)} - u_{(n-s+i)}]^{f_{ei}} \right\} [u_{(n-s+1)}^{\gamma} - u_{(r)}^{\gamma}]^{n-r-s}$$

$$\left\{ \prod_{i=1}^{r} \gamma u_{(i)}^{\gamma-1} \right\} \left\{ \prod_{i=1}^{s} \gamma u_{(n-s+i)}^{\gamma-1} \right\} d\boldsymbol{u},$$

$$0 < u_{(1)} < u_{(2)} < \cdots < u_{(r)} < u_{(n-s+1)} < \cdots < u_{(n)} < 1.$$

$$P_a = K\gamma^{r+s} \int \left\{ \prod_{i=1}^{r} [u_{(i)} - u_{(i-1)}]^{f_{pi}} \right\} [u_{(n-s+1)} - u_{(r)}]^{m-\sum_{i=1}^{r} f_{pi} - \sum_{i=1}^{s} f_{ei}} \left\{ \prod_{i=1}^{s} [u_{(n-s+i+1)} - u_{(n-s+i)}]^{f_{ei}} \right\}$$

$$\left[ \sum_{l=0}^{n-r-s} \binom{n-r-s}{l} (-1)^l u_{(n-s+1)}^{\gamma(n-r-s-l)} u_{(r)}^{\gamma l} \right] \left\{ \prod_{i=1}^{r} \gamma u_{(i)}^{\gamma-1} \right\} \left\{ \prod_{i=1}^{s} \gamma u_{(n-s+i)}^{\gamma-1} \right\} d\boldsymbol{u}, \text{ where } u_{(0)} = 0, u_{(n+1)} = 1.$$

Then we have

$$P_a = K\gamma^{r+s} \sum_{l=0}^{n-r-s} \binom{n-r-s}{l} (-1)^l$$

$$\left\{ \int \left\{ \prod_{i=1}^{r} [u_{(i)} - u_{(i-1)}]^{f_{pi}} \right\} [u_{(n-s+1)} - u_{(r)}]^{m-\sum_{i=1}^{r} f_{pi} - \sum_{i=1}^{s} f_{ei}} \left\{ \prod_{i=1}^{s} [u_{(n-s+i+1)} - u_{(n-s+i)}]^{f_{ei}} \right\} \left\{ \prod_{i=1}^{r-1} u_{(i)}^{\gamma-1} \right\} \left\{ \prod_{i=2}^{s} u_{(n-s+i)}^{\gamma-1} \right\} u_{(r)}^{\gamma(l+1)-1} u_{(n-s+1)}^{\gamma(n-r-s-l+1)-1} d\boldsymbol{u} \right\}.$$

Suppose,

$$\mathcal{D}_l = \left\{ \int \left\{ \prod_{i=1}^{r} [u_{(i)} - u_{(i-1)}]^{f_{pi}} \right\} [u_{(n-s+1)} - u_{(r)}]^{m-\sum_{i=1}^{r} f_{pi} - \sum_{i=1}^{s} f_{ei}} \left\{ \prod_{i=1}^{s} [u_{(n-s+i+1)} - u_{(n-s+i)}]^{f_{ei}} \right\} \left\{ \prod_{i=1}^{r-1} u_{(i)}^{\gamma-1} \right\} \left\{ \prod_{i=2}^{s} u_{(n-s+i)}^{\gamma-1} \right\} u_{(r)}^{\gamma(l+1)-1} u_{(n-s+1)}^{\gamma(n-r-s-l+1)-1} d\boldsymbol{u} \right\}.$$

Note that,

$$\int_0^{u_{(2)}} u_{(1)}^{f_{p1}+\gamma-1}\left(u_{(2)} - u_{(1)}\right)^{f_{p2}} du_{(1)} = u_{(2)}^{f_{p1}+f_{p2}+\gamma} \int_0^1 v^{f_{p1}+\gamma-1}(1-v)^{f_{p2}} dv.$$

Proceeding similarly, we get,

$\mathcal{D}_l =$

$B(f_{p1} + \gamma, f_{p2} + 1)\, B(f_{p1} + f_{p2} + 2\gamma, f_{p3} + 1) \ldots B\left(\sum_{i=1}^{r-1} f_{pi} + (r-1)\gamma,\ f_{pr} + 1\right)$

$B(\sum_{i=1}^{r} f_{pi} + r\gamma + \gamma l,\ m - \sum_{i=1}^{r} f_{pi} - \sum_{i=1}^{s} f_{ei} + 1)$

$$\int u_{(n-s+1)}^{m-\sum_1^s f_{ei}+\gamma(m-s)+\gamma-1} \left\{\prod_{i=1}^{s}\left[u_{(n-s+i+1)} - u_{(n-s+i)}\right]^{f_{ei}}\right\}\left\{\prod_{i=2}^{s} u_{(n-s+i)}^{\gamma-1}\right\} d\boldsymbol{u}.$$

$$= \mathcal{A}_l \int u_{(n-s+1)}^{m-\sum_1^s f_{ei}+\gamma(n-s)+\gamma-1} \left\{\prod_{i=1}^{s}\left[u_{(n-s+i+1)} - u_{(n-s+i)}\right]^{f_{ei}}\right\}\left\{\prod_{i=2}^{s} u_{(n-s+i)}^{\gamma-1}\right\} d\boldsymbol{u}.$$

where

$\mathcal{A}_l = B(f_{p1} + \gamma, f_{p2} + 1)\, B(f_{p1} + f_{p2} + 2\gamma, f_{p3} + 1) \ldots B\left(\sum_{i=1}^{r-1} f_{pi} + (r-1)\gamma,\ f_{pr} + 1\right)$

$B(\sum_{i=1}^{r} f_{pi} + r\gamma + \gamma l,\ m - \sum_{i=1}^{r} f_{pi} - \sum_{i=1}^{s} f_{ei} + 1).$

Let

$$\mathcal{B}_l = \int u_{(n-s+1)}^{m-\sum_1^s f_{ei}+\gamma(n-s)+\gamma-1} \left\{\prod_{i=1}^{s}\left[u_{(n-s+i+1)} - u_{(n-s+i)}\right]^{f_{ei}}\right\}\left\{\prod_{i=2}^{s} u_{(n-s+i)}^{\gamma-1}\right\} d\boldsymbol{u}.$$

Carrying out the integration for the remaining part in $\mathcal{D}$, we get

$\mathcal{B}_l = B(m + \gamma(n-s) - \sum_1^s f_{ei} + \gamma, f_{e1} + 1)\, B(m + \gamma(n-s) - \sum_2^s f_{ei} + 2\gamma, f_{e1} + 1) \ldots$

$B(m + \gamma(n-s) - \sum_{s-1}^{s} f_{ei} + (s-1)\gamma, f_{e(s-1)} + 1)\, B(m + \gamma(n-s) - f_{es} + s\gamma, f_{es} + 1).$

Therefore, under $H_{R1}$, we have $\mathcal{D}_l = \mathcal{A}_l \mathcal{B}_l$ and,

$P[f_{p1}, f_{p2}, \ldots, f_{pr}, \ldots, f_{e1}, f_{e2}, \ldots, f_{es} \mid H_1]$

$$= K\gamma^{r+s} \sum_{l=0}^{n-r-s} \binom{n-r-s}{l} (-1)^l \mathcal{D}_l$$

$$= K\gamma^{r+s} \sum_{l=0}^{n-r-s} \binom{n-r-s}{l} B(f_{p1}+\gamma, f_{p2}+1) \, B(f_{p1}+f_{p2}+2\gamma, f_{p3}+1) \ldots B\left(\sum_{i=1}^{r-1} f_{pi} + (r-1)\gamma, \, f_{pr}+1\right) \, B\left(\sum_{i=1}^{r} f_{pi} + r\gamma + \gamma l, \, m - \sum_{i=1}^{r} f_{pi} - \sum_{i=1}^{s} f_{ei} + 1\right) \, B(m+\gamma(n-s) - \sum_{1}^{s} f_{ei} + \gamma, f_{e1}+1) \, B(m+\gamma(n-s) - \sum_{2}^{s} f_{ei} + 2\gamma, f_{e1}+1) \ldots$$

$$B\left(m+\gamma(n-s) - \sum_{s-1}^{s} f_{ei} + (s-1)\gamma, \, f_{e(s-1)}+1\right) B(m+\gamma(n-s) - f_{es} + s\gamma, f_{es}+1).$$

The proof of Theorem 4.1 is completed.

## A2. Proof of Theorem 3.1.

We obtain the joint probability mass function of $\{f_{p1}, f_{p2}, \ldots, f_{pr}, \, f_{e1}, f_{e2}, \ldots, f_{es}\}$ under $H_0$ by taking $\gamma = 1$ in $P[f_{p1}, f_{p2}, \ldots, f_{pr}, \, f_{e1}, f_{e2}, \ldots, f_{es} \mid H_1]$. Therefore, we have

$$P[f_{p1}, f_{p2}, \ldots, f_{pr}, \, f_{e1}, f_{e2}, \ldots, f_{es} \mid H_0] = K \sum_{l=0}^{n-r-s} \binom{n-r-s}{l} (-1)^l \, \mathcal{B}_l^1 \, \mathcal{B}_l^2,$$

where,

$$\mathcal{B}_l^1 = \frac{\Gamma(f_{p1}+1)\Gamma(f_{p2}+1)}{\Gamma(f_{p1}+f_{p2}+2)} \, \frac{\Gamma(f_{p1}+f_{p2}+2)\Gamma(f_{p3}+1)}{\Gamma(f_{p1}+f_{p2}+f_{p3}+3)} \ldots \frac{\Gamma\left(\sum_{i=1}^{r-1} f_{pi}+(r-1)\right)\Gamma(f_{pr}+1)}{\Gamma\left(\sum_{i=1}^{r} f_{pi}+r\right)}$$

$$\frac{\Gamma\left(\sum_{i=1}^{r} f_{pi}+r+l\right)\Gamma\left(m-\sum_{1}^{r} f_{pi}-\sum_{1}^{s} f_{ei}+1\right)}{\Gamma\left(m-\sum_{1}^{s} f_{ei}+r+l+1\right)},$$

and,

$$\mathcal{B}_l^2 = \frac{\Gamma(m+(n-s)-\sum_{1}^{s} f_{ei}+1)\Gamma(f_{e1}+1)}{\Gamma(m+(n-s)-\sum_{2}^{s} f_{ei}+2)} \, \frac{\Gamma(m+(n-s)-\sum_{2}^{s} f_{ei}+2)\Gamma(f_{e2}+1)}{\Gamma(m+(n-s)-\sum_{3}^{s} f_{ei}+3)} \ldots \frac{\Gamma(m+(n-s)-\sum_{s-1}^{s} f_{ei}+(s-1))\Gamma(f_{e(s-1)}+1)}{\Gamma(m+(n-s)-f_{es}+s)}$$

$$\frac{\Gamma(m+(n-s)-f_{es}+s)\Gamma(f_{es}+1)}{\Gamma(m+(n-s)+s+1)}.$$

After simplifying $\mathcal{B}_l^1$ and $\mathcal{B}_l^2$, we get

$$P[f_{p1}, f_{p2}, \ldots, f_{pr}, f_{e1}, f_{e2}, \ldots, f_{es} | H_0] = K \sum_{l=0}^{n-r-s} \binom{n-r-s}{l} (-1)^l \frac{\prod_1^r \Gamma(f_{pi}+1)}{\Gamma(\sum_1^r f_{pi}+r)}$$

$$\frac{\left(\Gamma(\sum_1^r f_{pi}+r+l)\, \Gamma(m-\sum_1^r f_{pi}-\sum_1^s f_{ei}+1)\right)}{\Gamma(m-\sum_1^s f_{ei}+r+l+1)} \frac{(\Gamma(m+(n-s)-\sum_1^s f_{ei}+1)\, \prod_1^s(f_{ei}+1))}{\Gamma(m+(n-s)+s+1)}.$$

We have $K = \frac{m!\, n!}{\prod_1^r f_{pi}!\, (m-\sum_1^r f_{pi}-\sum_1^s f_{ei})!\, \prod_1^s f_{ei}\, (n-r-s)!}$. Therefore,

$$P[f_{p1}, f_{p2}, \ldots, f_{pr}, f_{e1}, f_{e2}, \ldots, f_{es} | H_0]$$

$$= \frac{m!\, n!}{\prod_1^r f_{pi}!\, (m-\sum_1^r f_{pi}-\sum_1^s f_{ei})!\, \prod_1^s f_{ei}\, (n-r-s)!} \frac{(\prod_1^r f_{pi}!)\,(m-\sum_1^r f_{pi}-\sum_1^s f_{ei})!}{(\sum_1^r f_{pi}+r-1)!} \frac{(m+(n-s)-\sum_1^s f_{ei})!\, \prod_1^s f_{ei}}{(m+n)!}$$

$$\sum_{l=0}^{n-r-s} \binom{n-r-s}{l} (-1)^l \frac{\Gamma(\sum_1^r f_{pi}+r+l)}{\Gamma(m-\sum_1^s f_{ei}+r+l+1)}$$

$$= \frac{m!\, n!}{\prod_1^r f_{pi}!\, (m-\sum_1^r f_{pi}-\sum_1^s f_{ei})!\, \prod_1^s f_{ei}\, (n-r-s)!} \frac{(\prod_1^r f_{pi}!\, \prod_1^s f_{ei}!)}{(m+n)!} \left(m - \sum_1^r f_{pi} - \sum_1^s f_{ei}\right)!$$

$$\sum_{l=0}^{n-r-s} \binom{n-r-s}{l} (-1)^l \frac{(\sum_1^r f_{pi}+r+l-1)!\, (m+(n-s)-\sum_1^s f_{ei})!}{(m-\sum_1^s f_{ei}+r+l)!\, (\sum_1^r f_{pi}+r-1)!}$$

$$= \frac{m!\, n!}{\prod_1^r f_{pi}!\, (m-\sum_1^r f_{pi}-\sum_1^s f_{ei})!\, \prod_1^s f_{ei}\, (n-r-s)!} \frac{(\prod_1^r f_{pi}!\, \prod_1^s f_{ei}!)}{(m+n)!} \left(m - \sum_1^r f_{pi} - \sum_1^s f_{ei}\right)!$$

$$\sum_{l=0}^{n-r-s} (-1)^l\, (n-r-s)!\, \frac{(\sum_1^r f_{pi}+r+l-1)!}{(\sum_1^r f_{pi}+r-1)!\, l!} \frac{(m+(n-s)-\sum_1^s f_{ei})!}{(m-\sum_1^s f_{ei}+r+l)!\, (n-r-s-l)!}$$

$$= \frac{m!\, n!}{(m+n)!} \sum_{l=0}^{n-r-s} (-1)^l \binom{\sum_1^r f_{pi}+r+l-1}{l} \binom{(m+(n-s)-\sum_1^s f_{ei})}{(n-s-r-l)}.$$

Note that,

$$\sum_{l=0}^{n-r-s} (-1)^l \binom{\sum_1^r f_{pi}+r+l-1}{l} \binom{(m+(n-s)-\sum_1^s f_{ei})}{(n-s-r-l)} = \text{Coefficient of } Z^{b-r-s} \text{ in the}$$

product of $(1+z)^{-(\sum_1^r f_{pi}+r)}$ and $(1+z)^{m+(n-s)-\sum_1^s f_{ei}}$.

Therefore, we have

$$\sum_{l=0}^{n-r-s}(-1)^l \binom{\sum_1^r f_{pi} + r + l - 1}{l} \binom{(m + (n - s) - \sum_1^s f_{ei})}{(n - s - r - l)}$$

$$= \text{Coefficient of } Z^{n-r-s} \text{ in } (1 + z)^{m-\sum_1^s f_{ei}-\sum_1^r f_{pi}+n-r-s} = \binom{m - \sum_1^s f_{ei} - \sum_1^r f_{pi} + n - r - s}{n - r - s}.$$

Thus, we obtain the joint probability mass function of $M_r$ and $M_s$ under $H_0$

$$P[f_{p1}, f_{p2}, \ldots, f_{pr}, f_{e1}, f_{e2}, \ldots, f_{es} \mid H_0]$$

$$= \frac{\binom{m - \sum_1^s f_{ei} - \sum_1^r f_{pi} + n - r - s}{n - r - s}}{\binom{m+n}{n}}.$$

The proof of Theorem 3.1 is completed.

### A3. Proof of Theorem 3.2.

We denote $\sum_1^r f_{pi} = N_1$, $\sum_1^s f_{ei} = N_2$, and $N = N_1 + N_2$. The problem of evaluating the sum

$$\sum_{\substack{f_{p1}, f_{p2}, \ldots, f_{pr}, f_{e1}, f_{e2}, \ldots, f_{es} \\ P_r = i \; ; \; E_s = k-i}} \frac{\binom{m - \sum_1^s f_{ei} - \sum_1^r f_{pi} + n - r - s}{n - r - s}}{\binom{m+n}{n}}$$ is same as distributing $N$ indistinguishable balls

in $(r + s)$ distinguishable boxes, by combinatorial argument. The balls are distributed in a way that maximum number of balls in a box from first $r$ boxes is $P_r = i$ and the maximum number of balls in a box from the last $s$ boxes is $E_s = j$ with $i + j = k$. Minimum number of balls in all the $(r + s)$ boxes is 0. Note that, $0 \leq P_r, E_s \leq k \leq N \leq m$, and $N_1 \leq rP_r$, $N_2 \leq sE_s$.

We have $N_1$ balls in $r$ boxes with $i$ maximum and 0 minimum to each box. Then $N_1$ balls can be distributed in $w(N_1, r, i, 0)$ ways that is obtained by the recursive formula given by

$$w(N_1, r, i, 0) = \sum_{c=0}^{i} \binom{N_1}{c} w((N_1 - c), (r - 1), i, 0), \tag{*}$$

where $w(0, 0, i, 0) = 1$ and $w(N_1, 0, i, 0) = 0$ when $N_1 \neq 1$.

Similarly, we have $N_2 = (N - N_1)$ balls in $s$ boxes with $(k - i)$ maximum and 0 minimum to each box. Then $N_2$ balls can be distributed in $w(N_2, s, (k - i), 0)$ ways that is obtained by the recursive formula given by

$$w(N_2, s, (k - i), 0) = \sum_{c=0}^{(k-i)} \binom{N_2}{c} w((N_2 - c), (s - 1), (k - i), 0) \qquad (**)$$

where $w(0, 0, (k - i), 0) = 1$ and $w(N_2, 0, (k - i), 0) = 0$ when $N_2 \neq 1$.

For a given $k$, $N$, and $P_r = i$, the probability $P[P_r = i, E_s = k - i]$ can be written as

$$P[P_r = i, E_s = k - i] = \sum_{\substack{f_{p1}, f_{p2}, \ldots, f_{pr}, \ldots, f_{e1}, f_{e2}, \ldots, f_{es} \\ P_r = i \; ; \; E_s = k - i}} \frac{\binom{m - \sum_1^s f_{ei} - \sum_1^r f_{pi} + n - r - s}{n - r - s}}{\binom{m+n}{n}}$$

$$= \sum_{N=0}^{\min(m, ri + s(k-i))} \sum_{N_1=0}^{\min(N, ri)} I((N - N_1) \leq s(k - i)) w(N_1, r, i, 0) \, w(N_2, s, (k - i), 0) \frac{\binom{m - N + n - r - s}{n - r - s}}{\binom{m+n}{n}},$$

where $I(.)$ is the standard indicator function. Therefore, the null distribution of $T_{r,s} = P_r + E_s$ is given by

$$P[T_{r,s} \leq t | H_0]$$

$$= \sum_{k=0}^{t} \sum_{i=0}^{k} P[P_r = i, E_s = k - i]$$

$$= \sum_{k=0}^{t} \sum_{i=0}^{k} \sum_{N=0}^{\min(m, ri + s(k-i))} \sum_{N_1=0}^{\min(N, ri)} I((N - N_1) \leq s(k - i)) w(N_1, r, i, 0) \, w((N - N_1), s, (k - i), 0) \frac{\binom{m - N + n - r - s}{n - r - s}}{\binom{m+n}{n}}$$

## A4. Proof of Theorem 3.3.

We derive the asymptotic distribution of $T_{r,s}$ under $H_0$. From Theorem 3.1., We have

$$P[f_{p1}, f_{p2}, \ldots, f_{pr}, f_{e1}, f_{e2}, \ldots, f_{es} \mid H_0] = \frac{\binom{m - \sum_1^s f_{ei} - \sum_1^r f_{pi} + n - r - s}{n - r - s}}{\binom{m+n}{n}}.$$

By expanding the factorials,

$$\frac{\binom{m - \sum_1^s f_{ei} - \sum_1^r f_{pi} + n - r - s}{n - r - s}}{\binom{m+n}{n}} = \frac{n(n-1)\ldots(n-r-s+1)\, m(m-1)(m-2)\ldots(m - \sum_1^s f_{ei} - \sum_1^r f_{pi} + 1)}{(m+n)(m+n-1)\ldots(m+n - \sum_1^s f_{ei} - \sum_1^r f_{pi} - r - s + 1)!}$$

$$= \frac{n^{r+s}\, m^{\sum_1^s f_{ei} + \sum_1^r f_{pi}} \left(1 - \frac{1}{n}\right)\left(1 - \frac{2}{n}\right)\ldots\left(1 - \frac{r+s-1}{n}\right)\left(1 - \frac{1}{m}\right)\left(1 - \frac{2}{m}\right)\ldots\left(1 - \frac{\sum_1^s f_{ei} + \sum_1^r f_{pi} - 1}{m}\right)}{(m+n)^{\sum_1^s f_{ei} + \sum_1^r f_{pi} + r + s} \left(1 - \frac{1}{m+n}\right)\ldots\left(1 - \frac{\sum_1^s f_{ei} + \sum_1^r f_{pi} + r + s}{m+n}\right)}.$$

With $m, n \to \infty$, and $\frac{m}{n} \to 1$, we get

$$\lim_{m,n \to \infty} \frac{\binom{m - \sum_1^s f_{ei} - \sum_1^r f_{pi} + n - r - s}{n - r - s}}{\binom{m+n}{n}} \approx \left(\frac{1}{2}\right)^{\sum_1^s f_{ei} + \sum_1^r f_{pi} + r + s}.$$

The proof of Theorem 3.3 is completed.

## References

Balakrishnan, N., and R. Frattina. 2000. Precedence test and maximal precedence test. *Recent Advances in Reliability Theory* (pp. 355-378). Birkhäuser, Boston, MA.

Balakrishnan, N., and H. K. T. Ng. 2006. *Precedence-type tests and applications*. Hoboken, New Jersey: John Wiley & Sons.

Balakrishnan, N., A. Dembińska, and A. Stepanov. 2008a. Precedence-type tests based on record values. *Metrika* 68(2): 233-255.

Balakrishnan, N., T. Li, and J. Zhang. 2021. Application of RSS design with rank-sum precedence tests for early decision on equality of two distributions. *Journal of Statistical Computation and Simulation* 91(4): 667-693.

Balakrishnan, N., C. Paroissin, and J. Turlot. 2015. One-sided control charts based on precedence and weighted precedence statistics. *Quality and Reliability Engineering International* 31(1): 113-134.

Balakrishnan, N., R. C. Tripathi, and N. Kannan. 2008b. On the joint distribution of placement statistics under progressive censoring and applications to precedence test. *Journal of Statistical Planning and Inference* 138(5): 1314-1324.

Balakrishnan, N., R. C. Tripathi, N. Kannan, and H. K. T. Ng. 2010. Some nonparametric precedence-type tests based on progressively censored samples and evaluation of power. *Journal of statistical planning and inference* 140(2): 559-573.

Chakraborty, N., S. W. Human, and N. Balakrishnan. 2018. A generally weighted moving average exceedance chart. *Journal of Statistical Computation and Simulation* 88(9): 1759-1781.

Erem, A. 2020. Bivariate two sample test based on exceedance statistics. *Communications in Statistics-Simulation and Computation* 49(9): 2389-2401.

Erem, A. and I. Bayramoglu. 2017. Exact and asymptotic distributions of exceedance statistics for bivariate random sequences. *Statistics & Probability Letters* 125:181-188.

Graham, M. A., S. Chakraborti, and A. Mukherjee. 2014. Design and implementation of CUSUM exceedance control charts for unknown location. *International Journal of Production Research* 52(18):5546-5564.

Graham, M. A., A. Mukherjee, and S. Chakraborti. 2017. Design and implementation issues for a class of distribution-free Phase II EWMA exceedance control charts. *International Journal of Production Research* 55(8):2397-2430.

Lawless, J. F. 2011. *Statistical models and methods for lifetime data*. Hoboken, New Jersey: John Wiley & Sons.

Li, T., Balakrishnan, N., H. K. T. Ng, Y. Lu, and L. An. 2019. Precedence tests for equality of two distributions based on early failures of ranked set samples. *Journal of Statistical Computation and Simulation* 89(12):2328-2353.

Mukherjee, A., Y. Cheng, and M. Gong. 2018. A new nonparametric scheme for simultaneous monitoring of bivariate processes and its application in monitoring service quality. *Quality Technology & Quantitative Management* 15(1):143-156.

Ng, H. K. T., and N. Balakrishnan. 2005. Weighted precedence and maximal precedence tests and an extension to progressive censoring. *Journal of Statistical Planning and Inference* 135(1):197-221.

Ng, H. K. T., and N. Balakrishnan. 2010. Precedence-type test based on Kaplan–Meier estimator of cumulative distribution function. *Journal of statistical planning and inference* 140(8):2295-2311.

Ng, H. K. T., N. Balakrishnan, and S. Panchapakesan. 2007. Selecting the best population using a test for equality based on minimal Wilcoxon rank-sum precedence statistic. *Methodology and Computing in Applied Probability* 9(2):263-305.

Seidler, J., J. Vondráček, and I. Saxl. 2000. The life and work of Zbyněk Šidák (1933–1999). *Applications of Mathematics 45*(5):321-336.

Šidák, Z., and J. Vondráček. 1957. A simple nonparametric test of the difference of location of two populations. *Apl. mat*, *2*, 215-221.

Stoimenova, E., and N. Balakrishnan. 2011. A class of exceedance-type statistics for the two-sample problem. *Journal of statistical planning and inference* 141(9):3244-3255.

Stoimenova, E., and N. Balakrishnan. 2017. Šidák-type tests for the two-sample problem based on precedence and exceedance statistics. *Statistics* 51(2):247-264.

van der Laan, P., and S. Chakraborti. 2001. Precedence tests and Lehmann alternatives. *Statistical Papers* 42(3):301-312.